\documentclass[twocolumn,prb,showpacs]{revtex4}
\usepackage[dvips]{graphics}
\usepackage[dvips]{color}
\begin{document}
\title{Reduction of the sign problem using the meron-cluster approach.}
\author{Sara Bergkvist}
\email{sara@theophys.kth.se}
\author{Patrik Henelius}
\author{Anders Rosengren}
\affiliation{Condensed Matter Theory, Physics Department, KTH,
             SE-106 91 Stockholm, Sweden}
\date{\today}

\begin{abstract}
The sign problem in quantum Monte Carlo calculations is analyzed using
the meron-cluster solution. The concept of merons can be used to solve
the sign problem for a limited class of models. Here we show that the
method can be used to \textit{reduce} the sign problem in a wider
class of models. We investigate how the meron solution evolves between
a point in parameter space where it eliminates the sign problem and a
point where it does not affect the sign problem at all. In this
intermediate regime the merons can be used to reduce the sign
problem. The average sign still decreases exponentially with system
size and inverse temperature but with a different prefactor. The sign
exhibits the slowest decrease in the vicinity of points where the
meron-cluster solution eliminates the sign problem. We have used
stochastic series expansion quantum Monte Carlo combined with the
concept of directed loops.
\end{abstract}
\maketitle

\section{Introduction} 

To stochastically study a quantum problem using a quantum Monte Carlo
method it is necessary to transform it to a form that is similar to a
classical statistical problem. The sign problem appears when this
transformation leads to a weight function that is not positive
definite. As quantum Monte Carlo methods have become increasingly
efficient\cite{ever93,kawa94,prok96,bear96,SaPRB99,SyPRE02} there is a
notable lack of progress in solving the sign problem. The sign problem
severely limits the number of models that can be studied using quantum
Monte Carlo methods, and in particular there are only very few models
of interacting fermions in higher dimensions that are accessible to
existing algorithms.\cite{blan81}

The recent development of the so-called meron-cluster
solution\cite{ChPRL99} has extended the range of models where the sign
problem can be avoided.  This method uses the properties of loop
quantum Monte Carlo algorithms to establish a one-to-one mapping
between configurations with negative weight and corresponding
configurations with positive weight. These contributions cancel each
other and a fraction of the phase space with a positive definite
weight function is left, which can be sampled with no sign problem.

Unfortunately the meron solution works for only a rather limited class
of models.\cite{Ch02} The main purpose of the present paper is to show
that the meron concept can be applied also to models where the sign
problem is not eliminated. We demonstrate that in a wider class of
models it is possible to cancel out part of the negative
configurations, and thereby reduce the sign problem. We investigate a
model in an intermediate regime between a point in parameter space
where the sign problem can be solved completely and a point where the
meron-cluster algorithm cannot be applied. Our results show that the
meron-cluster algorithm does indeed reduce the sign problem in this
intermediate regime. The main focus of the study is frustrated spin
models, but we also apply this method to spinless fermions.

The outline of the paper is as follows. In Sec.~\ref{sec2} the 
Monte Carlo algorithm is briefly explained. The sign problem and the
meron-cluster algorithm are introduced in Sec.~\ref{sec3}. In
Sec.~\ref{sec4} a modified version of the stochastic series expansion
is described. The origin of the sign problem for frustrated spin
systems and fermions is discussed in Sec.~\ref{sec5}.  In
Sec.~\ref{sec6} we demonstrate how the meron solution affects the
average sign for a range of models where the sign problem cannot be
eliminated. We conclude with summary and discussions in
Sec.~\ref{sec7}.

\section{The quantum Monte Carlo algorithm}
\label{sec2}
In order to explain the meron-cluster solution introduced in
Sec.~\ref{sec4} we here give a summary of the stochastic series
expansion (SSE) method\cite{sand91,SaPRB99,SyPRE02}.

Consider a lattice model described by a Hamiltonian $H$. In the SSE
method the partition function $Z$ is Taylor expanded,
\begin{equation}
Z=\sum_{\alpha}\sum_{m=0}^{\infty}\frac{\beta^m}{m!}
\langle\alpha|(-H)^m|\alpha\rangle,
\label{Z}
\end{equation}
where $ |\alpha\rangle$ are states in which the above matrix elements
can be calculated and $\beta$ denotes the inverse temperature.

For sake of clarity we will now consider a one-dimensional
ferromagnetic Heisenberg model,
\begin{equation}
H=-\sum_{i=1}^N S_i^zS_{i+1}^z + \frac{1}{2}\left(S_i^+S_{i+1}^-
+ S_i^-S_{i+1}^+\right),
\end{equation}
where $N$ denotes th number of sites. The Hamiltonian is rewritten as
a sum over diagonal and off-diagonal operators,
\begin{equation}
-H=\sum_{i=1}^N(H_{1,i}+H_{2,i}),
\label{Hdod}
\end{equation}
where 
\begin{equation}
H_{1,i}=S_{i}^zS_{i+1}^z+C
\end{equation}
 and
\begin{equation}
H_{2,i}=\frac{1}{2}(S_i^+S_{i+1}^-+S_i^-S_{i+1}^+),
\end{equation}
where $C$ is a constant inserted to assure that the expectation value
$\langle \alpha|H_{1,i}|\alpha\rangle $ is positive for all states $|
\alpha\rangle$.

To simplify the Monte Carlo update we introduce an additional unit
operator $H_{0,0}=1$.  Inserting the Hamiltonian given by
Eq.~(\ref{Hdod}) into Eq.~(\ref{Z}), and truncating the sum at $m=L$
we obtain
\begin{equation}
Z=\sum_{\alpha}\sum_{S_L}\frac{\beta^n(L-n)!}{L!}\langle
\alpha|\prod_{k=1}^L H_{a_k,i_k}|\alpha\rangle,
\label{Zeq}
\end{equation}
where $n$ stands for the number of non-unit operators, and $S_L$
denotes a sequence of operator-indices
\begin{equation}
S_L=(a_1,i_1),(a_2,i_2),...,(a_L,i_L),
\end{equation}
with $a_k={1,2}$ and $i_k={1,\ldots,N}$, or $(a_k,i_k)=(0,0)$. 

The Monte Carlo procedure must sample the space of all states
$|\alpha\rangle$, and all sequences $S_L$. The simulation starts with
some random state $|\alpha\rangle$ and an operator string containing
only unit operators. One Monte Carlo step consists of a diagonal and
an off-diagonal update.  In the diagonal update attempts are made to
exchange unit and diagonal operators sequentially at each position in
the operator string. The probability for inserting or deleting a
diagonal operator in the operator string is given by detailed
balance.\cite{SaPRB99}

\begin{figure}
\begin{center}
\resizebox{60mm}{!}{\includegraphics{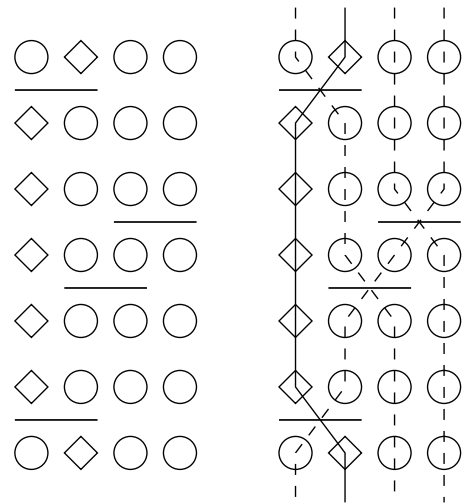}}
\end{center}
\caption{In the left part of the figure one of the terms in the
expansion in Eq.~(\ref{Zeq}) is shown. The two different states of the
basis, spin up and down, are illustrated with circles and diamonds and
the operators are depicted as horizontal bars. In the right part of
the figure the configuration is divided up into closed loops.  The two
loops in the configuration are distinguished by different line
styles.}
\label{cluster}
\end{figure}

The off-diagonal update, also called loop update, is carried out with
$n$ fixed.  Each bond operator $H_{i_k}=H_{1,i_k}+H_{2,i_k}$ acts only
on two spins, $S_{i_k}$ and $S_{i_k+1}$. We can therefore rewrite
the matrix elements in Eq.~(\ref{Zeq}) as a product of $n$ terms,
called vertices,
\begin{equation}
M(\alpha,S_L)=\prod_{k=1}^{n}W_k,
\label{vertw}
\end{equation}
where the vertex weight $W_p$ is defined as
\begin{eqnarray}
&&W_p=\langle S^z_{i_p}(p)S^z_{i_p+1}(p)|H_{i_p}|S^z_{i_p}(p-1),
S^z_{i_p+1}(k-1)\rangle\nonumber\\
\label{weight}
\end{eqnarray}
where $S^z_i(p)$ denotes the state of spin $i$ in a propagated
state, defined by
\begin{equation}
|\alpha(p)\rangle\sim\prod_{k=1}^{p}H_{a_k,i_k}|\alpha\rangle.
\end{equation}
A vertex thus consists of four spins, called the legs of the vertex,
and an operator. Each term in the expansion in Eq.~(\ref{Zeq}) can be
viewed as a sequence of vertices. An example of one term for a
four-site chain is shown in the left part of Fig.~\ref{cluster}.

The principles of the off-diagonal update are: one of the $n$ vertices
is chosen at random and one of its four legs is randomly selected as
the entrance leg. The spin of the entrance leg is flipped.  One of the
legs of the operator is chosen as the exit leg, and its state is also
changed. The exit leg is chosen with a probability calculated from the
weight of the obtained vertex.\cite{SaPRB99} Thereafter the vertex
list is sequentially searched for the next vertex that includes the
exit spin.  This spin becomes the entrance leg of the next vertex and
the procedure is continued until the original entrance leg is reached.
During one Monte Carlo step the loop update is repeated until on
average half of the vertices have been updated.

In the method described above the spin states are altered as the loop
is constructed. During one Monte Carlo step a given spin can be part
of several different loops, or it may be of none. In a few special
cases, such as for the isotropic Heisenberg model, the propagation of
each loop through the lattice is deterministic, meaning that there is
only one possible exit leg for each entrance leg.\cite{SaPRB99} In
these special cases it is possible to divide the whole space-time
lattice up into loops so that each and every spin belongs to only one
loop. An example of such a configuration is shown in the right part of
Fig.~\ref{cluster}.  The loop update can then be modified to
identifying the unique loop structure and flipping each loop with
probability one half.

\section{The Sign Problem}
\label{sec3}
In this section we show how the sign problem appears in quantum Monte
Carlo simulations and introduce the recent meron approach to solving
the sign problem. We start by considering a general form of an
expectation value that can be calculated by Monte Carlo methods,
\begin{equation}
 \langle A\rangle =\frac{\sum_i A(x_i)W(x_i)}{\sum_i W(x_i)}=\langle
 A(x)\rangle _W,
\label{expec}
\end{equation}
where the weight function $W(x_i)$ and $A(x_i)$ depend on the
configuration $x_i$.  When the coordinates $x_i$ are sampled according
to relative weight, the expectation value is given by the average
value of $A(x)$ as indicated in the last part of
Eq.~(\ref{expec}). The sign problem appears if the weight function is
not positive. In this case the sampling can be done using the absolute
value of the weight,
\begin{equation}
 \langle A\rangle=\frac{\langle As\rangle_{|W|}}{\langle s\rangle_{|W|}},
\label{expneg}
\end{equation}
where $s$ denotes the sign of the weight function and equals $\pm 1$.
However, in many cases of physical interest the average sign
approaches zero exponentially as the system size is increased. The
above expectation value will then suffer from very large statistical
fluctuations since it becomes a ratio of two small numbers. Let us now
consider how negative weight functions appear for quantum mechanical
systems. The weight function $W(\alpha, S_L)$ corresponding to the
partition function given by Eq.~(\ref{Zeq}) is
\begin{equation}
W(\alpha, S_L)=\frac{\beta^n(L-n)!}{L!}\langle
\alpha|\prod_{k=1}^L H_{a_k,i_k}|\alpha\rangle,
\end{equation}
This is strictly positive, and for the ferromagnet there is no sign
problem. Let us next consider an antiferromagnet. In this case the
diagonal and off-diagonal operators are of the form
\begin{equation}
H_{1,b}=-S_{i(b)}^zS_{j(b)}^z+C,
\end{equation}
 and
\begin{equation}
H_{2,b}=-\frac{1}{2}(S_{i(b)}^+S_{j(b)}^-+S_{i(b)}^-S_{j(b)}^+).
\end{equation}
By adjusting the constant $C$ it is still possible to have $\langle
\alpha|H_{1,b}|\alpha\rangle \ge 0$.  However, for the off-diagonal
operator the expectation value $\langle \alpha|H_{2,b}|\alpha\rangle
\le 0$, and the sign must be taken into account. If there is an odd
number of off-diagonal operators in the configuration the sign of the
weight function will be negative. Due to the periodic boundary
conditions in the imaginary time direction the number of off-diagonal
operators on a square lattice is always even and there is no sign
problem. However, if the system is frustrated, as on a triangular
lattice, the sign problem appears for the antiferromagnetic spin
model. For a fermionic system the anticommutator rules must be taken
into account and the sign of the configuration changes sign every time
two fermion world lines wrap around each other in imaginary time.  We
therefore see that both for frustrated spin models and fermionic
models the sign problem enters into the loop update. Flipping a loop
can cause the number of spin flipping operators to change parity, or
it can cause two fermions to permute and thereby change the
sign. Loops that cause the sign to change are called
merons,\cite{ChPRL99} and in some cases one can, in effect, solve the
sign problem by avoiding configurations that include merons.

In order to explain the meron solution we need to revisit the loop
update. As was pointed out at the end of the previous section it is
sometimes possible to divide the lattice up into a unique loop
structure. Expectation values can then be calculated using so-called
improved estimators, which are averages over all possible loop
configurations.  If the number of loops in the system is given by
$N_L$ there are $2^{N_L}$ configurations that can be reached by
flipping the loops, since all the loops can be in two states.  The
expectation value in Eq.~(\ref{expneg}) can therefore be rewritten as
\begin{equation}
\langle A\rangle=\frac{\langle \langle
A(x)s(x)\rangle\rangle_{|W|}}{\langle \langle s(x)\rangle
\rangle_{|W|}},
\label{expectExpand}
\end{equation}
where the double expectation brackets denote an average over the
different loop configurations
\begin{equation}
\langle \langle s(x)A(x)\rangle \rangle_{|W|}=\langle
\frac{1}{2^{N_L}}\sum_{l=1}^{2^{N_L}}s(x_l)A(x_l)\rangle_{|W|}.
\end{equation}

If certain criteria are fulfilled, the expectation value can be
expressed as
\begin{equation}
\langle A\rangle=\langle \langle
A(x)\delta_{n_M,0}\rangle\rangle_{|W|},
\label{expanded}
\end{equation}
where $n_M$ is the number of merons. Therefore only configurations
without sign changing loops give non-zero contributions to the
expectation value, and the sign problem is, in effect, solved.

Let us examine the necessary conditions for this to be the case:

\begin{enumerate} 
\item The lattice can be divided up into loops so that each spin
belongs to one and only one loop.

\item The weight must not change when the loops are flipped.

\item The loops must affect the sign independently.

\item The zero-meron sector must be positive definite.

\item The expectation value of the operator is unchanged when a loop
is flipped.
\end{enumerate}

Together these conditions place severe restrictions on which models
can be studied with the meron solution. Our aim is to examine if the
conditions can be relaxed to allow for a more general algorithm. Of
the five conditions the last one is the least severe. Many operators
for which this condition does not hold can be expressed by introducing
the two-meron sector.\cite{ChPRL99,HePRB00} Examples of operators for
which the last condition holds are the energy and heat capacity, which
in SSE are given by
\begin{eqnarray}
E&=&-\frac{\langle n\rangle}{\beta} \\
C&=& \langle n^2\rangle- \langle n\rangle^2 - \langle n\rangle.
\end{eqnarray}
where $n$ is the number of non-unit operators in the operator string. 

The first three conditions are severe and we are not aware of a way to
relax these restrictions. In this study we examine the case when the
fourth condition is not met, and the zero-meron sector is not positive
definite. If this is not the case Eq.~(\ref{expanded}) must be
replaced by
\begin{equation}
\langle A\rangle=\frac{\langle \langle A(x)s(x)\delta_{n_M,0}
\rangle\rangle_{|W|}}{\langle \langle s(x)\delta_{n_M,0}\rangle
\rangle_{|W|}},
\label{zero_weight}
\end{equation}
where $s(x)$ is the sign of the configuration in the zero meron
sector.  The higher meron sectors do not contribute to the average and
the sign problem is therefore reduced, but not eliminated.  To find
out how the average sign changes by leaving out the non-zero meron
sectors is the aim of this investigation. It is not clear whether the
exponential character of the sign problem will change, or how the
system size will affect the average sign. This depends on the relative
weight of the different meron sectors and on the average sign in the
zero-meron sector.

In order to investigate systems where condition four does not hold it
is necessary to satisfy the first three criteria.  As was stated at
the end of Sec.~\ref{sec2} there are a few special cases where a
unique loop structure exists, and in these cases the first condition
is automatically satisfied. For more general models, where one can
choose between different exit legs, this is not true, since there no
longer exists a \textit{unique} way to divide the lattice into
loops. It is, however, possible to cover the lattice with loops so
that each spin belongs to one and only one loop. This can be done by
always choosing an exit leg that is not already part of any
loop. Since a unique loop structure does not exist one also has to
sample all possible loop configurations, and in order to implement the
meron solution efficiently this has to be done without leaving the
zero and two-meron sectors. We have implemented a loop update that
inherently divides the lattice up in separate loops. This can be done
as long as the bounce process, where the entrance and exit legs
coincide, can be neglected. With the advent of directed
loops\cite{SyPRE02} there now exists a way to eliminate the bounce
process in many models of interest. In the next section we will
introduce the idea of directed loops and intoroduce models where the
first three, but not the forth, conditions are satisfied.
\section{SSE Loop Construction}
\label{sec4}

Following Ref.~\onlinecite{SyPRE02} we here derive the
directed loops equations that enable us to neglect the bounce process,
and explicitly show that the weight in the extended space of directed
loops is unaltered when a loop is flipped. Then we will describe a
modification to the SSE loop update and demonstrate how the meron
approach can be efficiently implemented also for models where there is
not a unique way of dividing the lattice into loops.

By considering that each possible loop has a ``time-reversed''
counterpart it was shown\cite{SyPRE02} that a loop move
satisfies detailed balance if, for every vertex that the loop passes
through,
\begin{equation}
W_sP(s,i\rightarrow j)=W_{s'}P(s',j\rightarrow i),
\label{det_bal}
\end{equation}
where the vertex weight $W_s$ for a given spin configuration $s$ is
defined by Eq.~(\ref{weight}) and $P(s,i\rightarrow j)$ is the
probability to choose exit leg $j$ given entrance leg $i$ for a given
spin configuration $s$.

The main idea of the method of directed loops is to attach weight also
to the link that connects the entrance and exit spins at a traversed
vertex. The weight of a vertex in this extended space can then be
written as $W(s,i,j)$. We are allowed to introduce such auxiliary
variables if the sum over auxiliary variables reduces to the original
weight
\begin{equation}
\sum_j W(s,i,j)=W_s. 
\label{sum}
\end{equation}
 We furthermore require that
the weight in this extended space is not affected by flipping a loop, which
translates to 
\begin{equation}
W(s,i,j)=W(s',j,i).
\label{nochange}
\end{equation}	
Considering the criteria for detailed balance, Eq.~(\ref{det_bal}) we
can now relate the weights in the original and extended space as
\begin{equation}
W(s,e,x)=W_sP(s,e,x).
\end{equation}

We have therefore shown that the total weight in our extended
configuration space is unchanged when flipping a loop, and the second
condition for using the meron solution is therefore fulfilled. The
method of directed loops is a way to assign probabilities to the
different possible exit legs given an entrance leg. The probabilities
are determined by solving the system of equations that
Eq.~(\ref{det_bal}) and Eq.~(\ref{nochange}) generate. We want to
emphasis that for many models it is possible to solve the equations so
that the bounce process is forbidden\cite{SyPRE02}, and as we now show
this makes it possible to divide the lattice up into separate loops.

Next we describe a new procedure that enables us to make this
division. In our new update the physical structure of the loops is
determined and updated during the diagonal move, while each loop is
flipped with probability one half in the off-diagonal update.  The
diagonal move is quite different from the one described in
section~\ref{sec2} and we here describe it in some detail. 

\begin{figure}
\begin{center}
\resizebox{80mm}{!}{\includegraphics{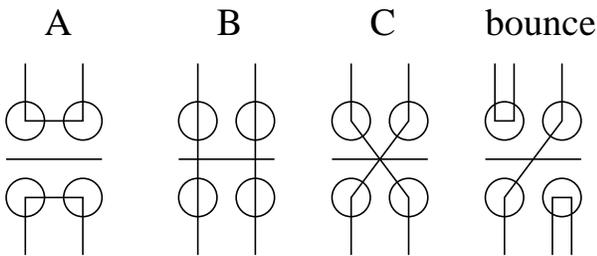}}
\end{center}
\caption{The three different ways the legs of a vertex can be paired
together, if the bounce process is neglected, are marked by letters
A,B and C. The fourth vertex shows one example of the many possible ways
to connect the legs if  the bounce is allowed. }
\label{struc}
\end{figure}

\begin{figure}
\begin{center}
\resizebox{60mm}{!}{\includegraphics{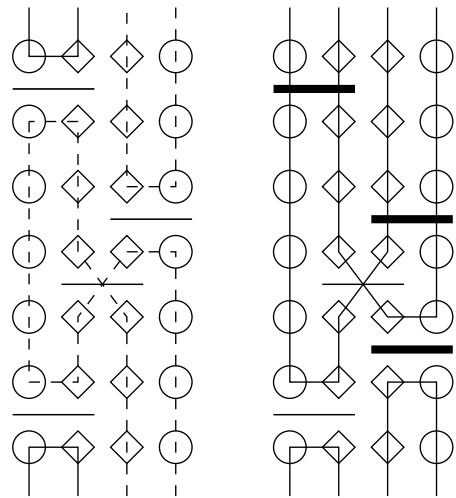}}
\end{center}
\caption{ A example of a SSE configuration divided up into separate
loops is depicted on the left side.  In the right part of the figure a
possible outcome after a sequential diagonal update is shown. One
operator has been inserted and the way the legs are connected has been
changed for two of the vertices. These operators are marked with a
thick bar.}
\label{changed} 
\end{figure}

The operator string is, as before, updated sequentially, and if a unit
operator is encountered an attempt to insert a diagonal operator is
made. If an operator is inserted its four legs are linked to the
existing loops.  If the bounce is eliminated there are three possible
ways to link the entrance and exit legs of a vertex, see
Fig.~\ref{struc}. The decision of which configuration to use is based
on detailed balance equations, which states that the probability of
different vertices is proportional to their weight given by
Eq.~\ref{weight}. It is enough to treat an arbitrary leg as ``entrance
leg'' to the vertex and determine an exit leg. With two legs connected
together in this manner the other two legs are also automatically
linked. If instead a diagonal operator is encountered, an attempt is
made to exchange it for a unit operator, as in the standard diagonal
update. If the removal is not accepted an attempt is made to alter the
way in which the legs are connected.  In the same way as when
inserting a new diagonal operator an ``exit leg'' is determined for an
arbitrary ``entrance leg'', a procedure which may change the existing
links . This change of loop structure is also done if an off-diagonal
operator is encountered. We graphically demonstrate this modified
update in Fig.~\ref{changed}. On the left side of Fig.~\ref{changed} a
term in the SSE expansion is divided up into loops. A possible outcome
after a diagonal update is shown on the right. In the next section we
will apply this update to models that suffer from the sign problem.

\section{Models}
\label{sec5}
\subsection{Frustrated spin}
\label{sec5a}

\begin{figure}
\begin{center}
\resizebox{25mm}{!}{\includegraphics{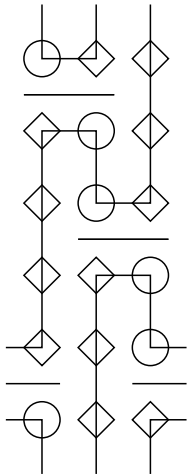}}
\end{center}
\caption{A three-spin configuration with negative weight but no
merons. The diamonds (circles) represent spin up (down).}
\label{negativ}
\end{figure}

The main focus of this article is frustrated spin systems. The
Hamiltonian, $H$, is given by,
\begin{equation}
H=\sum_{i,j}\Delta S^z_iS^z_{j}+\frac{1}{2}(S_i^+S_{j}^-+S_{j}^+S_{i}^-),
\label{spins}
\end{equation}
where the sum runs over all nearest neighbors. For $\Delta\in\{-1,1\}$
the directed loops equations can be solved so that the bounce is
eliminated, and using the method described in the previous section the
first two criteria for using the meron solution are fulfilled. The
frustration is introduced by having spins positioned on a triangular
lattice. The off-diagonal coupling is antiferromagnetic and as
described previously the number of off-diagonal operators may be odd
on a frustrated lattice and there is therefore a sign problem.  In
Fig.~\ref{negativ} an example of such a configuration with an odd
number of off-diagonal operators is shown. The third criteria states
that the loops need to be independent in their effect on the sign.  If
flipping a loop causes the total number of off-diagonal operators to
change parity the sign changes and the loop is a meron. The only way
this can happen is if the loop traverses an odd number of
vertices. The number of transversed vertices is independent of any
other loops and it is therefore clear that the loops are independent
in their effect on the sign, which establishes the third criteria. We
have therefore shown that the first three criteria for the application
of the meron solution are fulfilled.

The fourth criteria requires that the zero-meron sector has a positive
definite weight.  This is not, in general, true for this model. The
configuration shown in Fig.~\ref{negativ} has negative weight, yet the
only loop in the system is not a meron.  The only point where the
fourth criteria hold true is for $\Delta=-1$, a point in parameter
space which has been extensively studied.\cite{HePRB00} At this point
it is possible to exclude all but one update, update C in
Fig.~\ref{struc}. As we move away from $\Delta=-1$ the negative part
of the zero-meron sector grows. At the other extreme, $\Delta=1$, it
is not possible to eliminate the bounces without introducing a
non-ergodicity. If the bounce is eliminated the only allowed update is
update A in Fig~\ref{struc}. With only this update there are no loops
which pass through an odd number of operators, and thus there are no
merons. Still the zero-meron sector, the only sector left, contains
both positive and negative parts, but there is no way to switch
between them and therefore the whole SSE space is not sampled. For all
other values $\Delta\in\{-1,1\}$ the meron solution can be applied,
and this model constitutes an ideal testing ground for a further study
of the meron solution since, by adjusting a single parameter, we can
move from a point where the meron solution eliminates the sign problem
($\Delta=-1$) to a point where the loop update becomes non-ergodic
($\Delta=+1$). However, since much effort has been made to solve the
sign problem in fermion models, we will next describe the application
to a system of spinless fermions.

\subsection{Spinless fermions}
\label{sec5b}
Besides the frustrated spin systems we have also studied spinless
fermions on a square lattice. In this context a meron is a loop which
permutes an even number of fermions when it is flipped. The loops must
affect the sign independently of each other for the meron
method to work, as stated by the third constraint described in
Sec.~\ref{sec3}. For the frustrated spin models described above this
was always the case, but it is not so for fermionic models. As
discussed previously there are three different ways to traverse a
vertex, see Fig.~\ref{struc}. For fermions update C make the loops
dependent of each other.\cite{Ch02} To solve this problem one can give
the completely empty vertex a negative weight, but in this case update
B causes the loops to be dependent. We are not aware of a way solve
this dependency problem and one is thus restricted to models where one
of the two updates B and C can be forbidden.

In the work by Chandrasekharan et. al\cite{Ch02}, two models are given
where only one type of vertex update is allowed and where the zero
meron sector, therefore, is positive. We have studied a model where
it is possible to exclude update B and the bounce but where both
update A and C must be allowed. The Hamiltonian for this model is
\begin{eqnarray}
H&=&\sum_{i,j}c_i^+c_j^-+c_j^+c_i^-+\frac{4}{3}(n_i-1/2)(n_j-1/2)\nonumber\\
&&-\frac{1}{3}(n_i+n_j),
\label{elec}
\end{eqnarray}
where $n_i$ is the occupation on site $i$ and $c_i^-$ ($c_i^+$) are
the ordinary annihilation (creation) fermion operators. By adjusting
the constant, $C$ added to the diagonal operator, the empty vertex is
given a negative weight and the loops are independent of each other.
Therefore we have a fermionic example of a model where the first three
criteria for using the meron solution are fulfilled, but where the
zero-meron sector is not positive definite. In the next section we
analyze how the meron solution affects the average sign in these cases
where the zero-meron sector is not positive definite.

\section{Results}
\label{sec6}
\begin{figure}
\begin{center}
\resizebox{60mm}{!}{\includegraphics{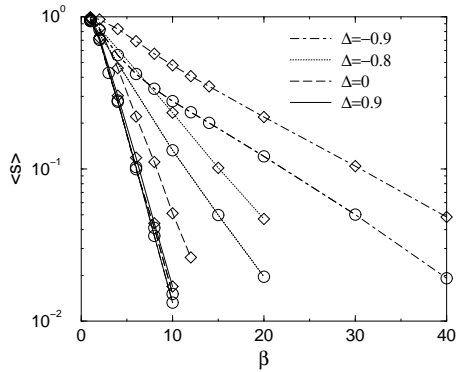}}
\end{center}
 \caption{The average sign for different values of $\Delta$ for the
frustrated spin model. The averages are calculated for the zero-meron
sector (diamonds) and including the two-meron sector (circles).}
\label{delta}
\end{figure}

First we consider the spin model described by Eq.~(\ref{spins}). We
have calculated the expectation value of the sign for frustrated spin
models with different values of the constant $\Delta$. The expectation
value for a three-site system is shown as a function of temperature in
Fig.~\ref{delta}. The simulation is performed in the zero and two
meron sectors and expectation values are shown both including and
excluding the two-meron sector.  As can be seen in the figure the sign
decreases exponentially in both cases. Asymptotically it appears that
the sign in the zero-meron sector is increased by a constant factor as
compared to the case of including the two-meron sector. The average
sign is greatest close to the point $\Delta=-1$, where the sign
problem is eliminated, and it decreases as the point $\Delta=1$ is
approached.  This is to be expected since the meron-cluster solution
cannot be applied in the present form at the Heisenberg
point($\Delta=1$).

\begin{figure}
\begin{center}
\resizebox{60mm}{!}{\includegraphics{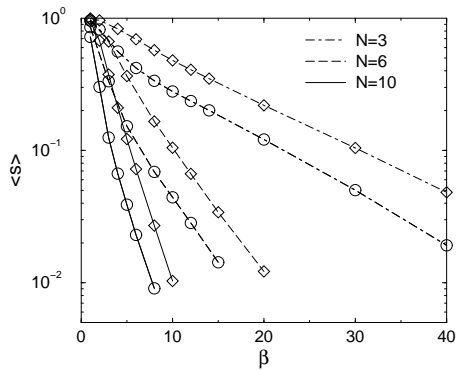}}
\end{center}
\caption{The average sign for different system sizes calculated for
$\Delta=-0.9$. The averages are calculated for the zero-meron sector
(diamonds) and including the two-meron sector (circles).}
\label{nbr}
\end{figure}

We have also calculated how the expectation value of the sign changes
with the system size. The results are shown in Fig.~\ref{nbr}. In this
calculation $\Delta=-0.9$ is used. The average sign appears to decrease
approximately exponentially also with system size.

\begin{figure}
 \begin{center}
   \resizebox{80mm}{!}{\includegraphics{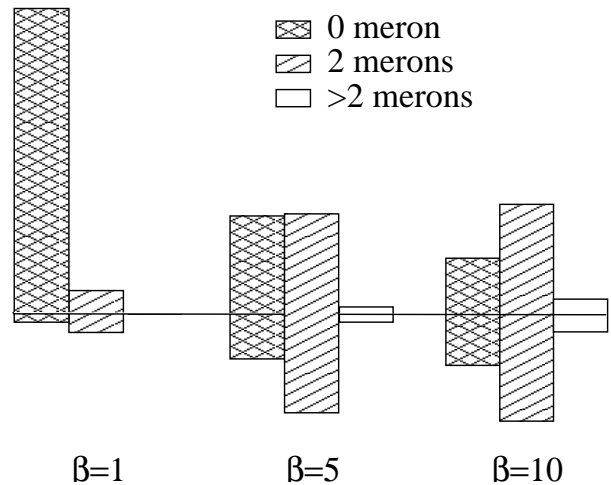}}
 \end{center}
\caption{The weight distribution between the sectors with different
number of merons at different temperatures for a system with ten
spins. The points above (below) the line in the figure correspond to a
positive (negative) sign.}
\label{dist}
\end{figure}

Besides the two already mentioned expectation values of the sign we
have also calculated the expectation value of the sign without the use
of merons. This value is slightly smaller than the one for the
two-meron sector. The difference is small since the sector with two
merons has a very much larger weight than the sector with more than
two merons, at least for the temperatures and system sizes that we
have studied. In Fig.~\ref{dist} we show the relative weight for the
zero-meron, the two-meron and the sector with more than two merons,
calculated for a system with ten spins at different temperatures.  The
weight of the sector with more than two merons increases with a
decreased temperature as a consequence of a larger configuration which
results in a larger number of loops.  In the figure we also indicate
the relation between the number of positive and negative values in the
different sectors. For the sectors with merons there is, by
definition, an equal amount of positive and negative values.  As can
be expected from Fig.~\ref{delta} the ratio of positive to negative
weight in the zero-meron sector approaches one as the temperature
decreases.  We also note that the relative weight of the zero-meron
sector decreases quite rapidly with lower temperatures. There is no
sector with an odd number of merons. If there where such a sector
configurations would exist where a flip of all the loops would result
in a sign change. A flip of all the loops corresponds to a change of
all spin states, an operation which does not  change the number of
off-diagonal operators.

\begin{figure}
\begin{center}
\resizebox{60mm}{!}{\includegraphics{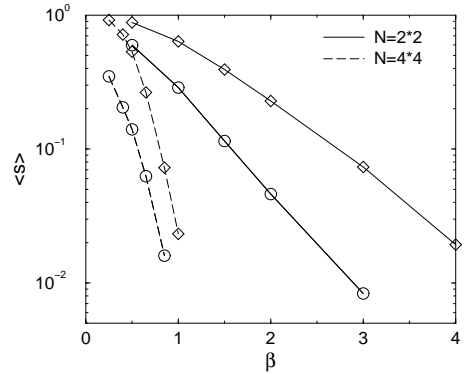}}
\end{center}
\caption{The average sign as a function of the temperature for the
fermion model. The averages are calculated for the zero-meron sector
(diamonds) and including the one- and two-meron sector (circles).}
\label{ferm}
\end{figure}

We have also done a similar calculation for the spinless fermion model
given by Eq.~(\ref{elec}). The result for the expectation value of the
sign is shown in Fig.~\ref{ferm} as a function of temperature. Two
different system sizes are studied, two times two and four times four
sites. Also here the average sign appears to decrease exponentially
with inverse temperature and system size both in the zero- and
two-meron sectors. Asymptotically it seems that the average sign in
the zero-meron sector again is increased by a constant factor when 
leaving out the higher meron sectors.

\begin{figure}
\begin{center}
\resizebox{80mm}{!}{\includegraphics{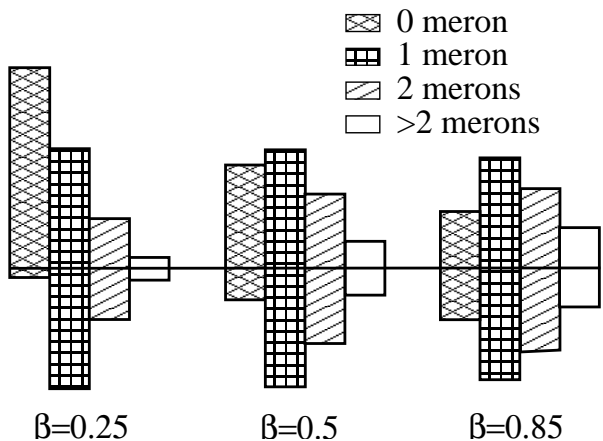}}
\end{center}
\caption{The weight distribution between the sectors with different number 
of merons at different temperatures for a system with
sixteen sites is presented. The points above (below) the line in the figure 
correspond to a positive (negative) sign.}
\label{distferm}
\end{figure}

We have studied the relative weight of the sectors with different
number of merons also for the fermionic system. For the fermions there
are configurations with only one meron. This is due to the fact that
the empty vertex is given a negative weight and flipping all the loops
in a configuration may change the parity of the number of empty
vertices.  A comparison with exact diagonalization indicates that one
only needs to include either the one- or the two-meron sector in
addition to the zero-meron sector.  In figure~\ref{distferm} the
relative weight for the zero-, one-, two-, and higher merons sectors
are presented.  At high temperatures the relative weight of the
zero-meron sector again dominates, while at lower temperatures the
weight in the higher meron sectors increases.

\section{Summary and discussion}
\label{sec7}
We have shown that as long as it is possible to divide the system into
independent loops, the meron-cluster approach can be used to decrease
the sign problem even when the zero-meron sector is not positive
definite. We have applied this method to both frustrated spin systems
and spinless fermions.  An intermediate regime between a point in
parameter space where the meron-cluster algorithm eliminates the sign
problem and a point where it cannot be applied is studied. In this
intermediate regime the exponential character of the sign problem
persists, but one can increase the average sign by a constant factor
by limiting measurements to the zero-meron sector. The method is
probably of most practical use in the vicinity of points where the
sign problem can be eliminated using the meron solution. In a large
scale application the weight of the two-meron sector should be
decreased by a reweighting technique\cite{HePRB00} to obtain better
statistics. To be able to use this algorithm we have combined
stochastic series expansion with the concept of directed loops.

\begin{acknowledgments}
We are grateful to A. Sandvik for stimulating discussions. The work
was supported by the Swedish Research Council and the G\"oran
Gustafsson foundation.
\end{acknowledgments}

\bibliography{bib} 
\end{document}